\documentclass[epj]{svjour}
\usepackage{graphicx}
\begin{document}
\title{Optical frequency measurement of the 1S-3S two-photon transition in hydrogen}
\author{O. Arnoult\inst{1} \and F. Nez\inst{1} \and L. Julien\inst{1} \and F. Biraben\inst{1}}

%
%
\institute{Laboratoire Kastler Brossel, \'{E}cole Normale Sup\'{e}rieure, CNRS, Universit\'{e} P. et M. Curie - Paris 6, Case 74, 4 place Jussieu, 75252 Paris Cedex 05, France}
\date{Received: date / Revised version: date}
%
\abstract{
This article reports the first optical frequency measurement of the $1\mathrm{S}-3\mathrm{S}$ transition in hydrogen. The excitation of this transition occurs at a wavelength of 205 nm which is obtained with two frequency doubling stages of a titanium sapphire laser at 820 nm. Its frequency is measured with an optical frequency comb. The second-order Doppler effect is evaluated from the observation of the motional Stark effect due to a transverse magnetic field perpendicular to the atomic beam. The measured value of the $1\mathrm{S}_{1/2}(F=1)-3\mathrm{S}_{1/2}(F=1)$ frequency splitting is $2~922~742~936.729~(13) \mathrm{MHz}$ with a relative uncertainty of $4.5\times10^{-12}$. After the measurement of the $1\mathrm{S}-2\mathrm{S}$ frequency, this result is the most precise of the optical frequencies in hydrogen.
} 
\maketitle
\section{Introduction}
\label{intro}
The hydrogen atom has a central position in the history of atomic physics. As it is the simplest of atoms, it has played a key role in testing fundamental theories, and hydrogen spectroscopy is associated with successive advances in the understanding of the atomic structure. Since the advent in the seventies of tunable lasers and methods of Doppler free spectroscopy, hydrogen spectroscopy has been renewed in the last decades. Consequently, several optical frequencies of hydrogen are now known with a fractional accuracy better than $10^{-11}$. In a long series of experiments, H\"{a}nsch and coworkers have improved the precision on the measurement of the $1\rm S-2\rm S$ frequency to obtain now a relative uncertainty of about $1.4\times 10^{-14}$ \cite{1S-2Sa}. In Paris, we have studied in the nineties the $2\mathrm{S}-n\mathrm{S/D}$ two-photon transitions with $n=8~\mathrm{and}~12$ ($n$ is the principal quantum number) \cite{{Ryd97},{Ryd99},{EPJD00}}. For instance we have measured the frequency of the $2\mathrm{S}_{1/2}-8\mathrm{D}_{5/2}$ transition with an uncertainty of $5.9~\mathrm{kHz}$, i.e. a relative uncertainty of $7.6\times 10^{-12}$. The goal of these high precision measurements is to determine the Rydberg constant $R_\infty$ and the hydrogen Lamb shifts.

The hydrogen energy levels can be conventionally expressed as the sum of three terms: the energy given by the Dirac equation for a particle with the reduced mass, the first relativistic correction due to the recoil of the proton and the Lamb shift. The first two terms have an exact expression as a function of the quantum numbers and of the fundamental constants (the Rydberg constant $R_\infty$, the fine structure constant  $\alpha$ and the electron-to-proton mass ratio $m_{e}/m_{p}$). The Lamb shift takes into account all the other corrections: corrections due to quantum electrodynamics (QED), other corrections due to the recoil of the proton and the effect of the proton charge distribution. The calculation of the Lamb shift is very difficult and a review of the results obtained so far is made in the report of the CODATA (Committee on Data for Science and Technology) \cite{codata06}. Today, for the $1\mathrm{S}$ level, the uncertainties are due to the calculation of the two-loop and three-loop QED corrections (this uncertainty is estimated to $3.7~\mathrm{kHz}$) and to the measurement of the proton charge distribution. Using the value of the radius $r_{\mathrm{p}}$ of this charge distribution deduced from the electron-proton scattering experiments ($r_{\mathrm{p}}=0.895\,(18)~\mathrm{fm}$) \cite{Sick}, this uncertainty is $50~\mathrm{kHz}$. At this level, the uncertainty of the theoretical value of the Lamb shift is not limited by the ones of $R_\infty$, $\alpha$  and $m_{e}/m_{p}$. Consequently, the uncertainty in the proton radius $r_{\mathrm{p}}$ is a strong limitation to extract the Rydberg constant from high precision measurements. For instance, using the frequency of the $1\rm S-2\rm S$ transition and the calculated value of the $1\mathrm{S}$ and $2\mathrm{S}$ Lamb shifts, the Rydberg constant can be deduced with a relative uncertainty of only  $1.8\times 10^{-11}$ \cite{LesHouches}.

In practice, it is possible to avoid this difficulty by using the $1/n^3$ scaling law for the Lamb shift. Numerous terms of the Lamb shift vary with the principal quantum number exactly as $1/n^{3}$ (for instance the effect of the charge distribution of the nucleus), and the deviation from this scaling law has been precisely calculated by Karshenboim \cite{Karshenboim}, and more recently by Pachucki \cite{Pachucki}. Then it is possible to eliminate the Lamb shift by forming a suitable linear combination of two frequency measurements. For instance, in the linear combination $7\nu(2\rm S_{1/2}-8\rm D_{5/2})-\nu(1\rm S_{1/2}-2\rm S_{1/2})$ of the  $2\mathrm{S}_{1/2}-8\mathrm{D}_{5/2}$ and $1\mathrm{S}_{1/2}-2\mathrm{S}_{1/2}$ frequencies, the quantity $L_{1\mathrm{S}}-8L_{2\mathrm{S}}$ appears and the effect of the proton charge distribution is eliminated ($L_{1\mathrm{S}}$ and $L_{2\mathrm{S}}$ are the Lamb shifts of the $1\mathrm{S}$ and $2\mathrm{S}$ levels). For example, from the measurements of the $1\rm S_{1/2}-2\rm S_{1/2}$, $2\rm S_{1/2}-8\rm D_{5/2}$ and $2\rm S_{1/2}-12\rm D_{5/2}$ frequencies in hydrogen and deuterium, one obtains a value of $R_{\infty}$ with a relative uncertainty of $9 \times 10^{-12}$ \cite{EPJD00}. Moreover, this method gives the values of the Lamb shifts, and, taking into account the theoretical calculations of the Lamb shift, it is possible to deduce a value of the rms charge radius of the proton ($r_{\mathrm{p}}=0.8746\,(94)~\mathrm{fm}$) which is more precise than the one deduced from the electron-proton scattering experiments. In this method, the accuracy is presently limited by the uncertainties of the $2\rm S_{1/2}-8\rm D_{5/2}$ and $2\rm S_{1/2}-12\rm D_{5/2}$ frequency measurements. In our experiment \cite{EPJD00}, these uncertainties were mainly constrained by the natural width of $8\mathrm{D}$ and $12\mathrm{D}$ levels and by the inhomogeneous light shift experienced by the atoms passing through the gaussian profile of the laser beams.

To circumvent the limitation due to the uncertainties of the $2\mathrm{S}-n\mathrm{D}$ measurements, our group studies the $1\rm S-3\rm S$ two-photon transition, with the aim to deduce $R_\infty$ and $r_{\mathrm{p}}$ from the comparison between the $1\rm S-2\rm S$ and $1\rm S-3\rm S$ frequencies. Indeed, as, in an atomic beam, the number of hydrogen atoms in the $1\mathrm{S}$ level is about eight orders of magnitude larger than the number of metastable atoms, the $1\rm S-3\rm S$ transition can be observed with very low light intensity so with a negligible light shift. In 1996, we have observed the $1\rm S-3\rm S$ transition and deduced, from the comparison with the $2\mathrm{S}-6\mathrm{S/D}$ frequency intervals, a value of the $1\mathrm{S}$ Lamb shift \cite{Bourzeix}. Since then we have undertaken optical frequency measurements of the $1\rm S-3\rm S$ transition. Because there is no simple way to determine the velocity distribution of the $1\rm S$ atomic beam, a difficulty is the determination of the second order Doppler effect, which induces a red shift of $-v^2/2c^2$. To measure it, the atomic beam is placed in a transverse magnetic field which induces a motional electric field and so a quadratic Stark shift varying also as $v^2$. The velocity of the atoms and the second order Doppler effect are deduced from the variation of this Stark shift with the magnetic field \cite{{Doppler},{Hagel}}. With these techniques, we have made a preliminary measurement of the $1\rm S-3\rm S$ optical frequency.

The aim of this paper is to relate in detail these last measurements. Section 2 describes the experimental method. Section 3 and 4 are devoted to the method used to determine the velocity distribution and to the theoretical analysis of the line shape. Finally the results are presented and analyzed in section 5.

\section{Spectroscopy of the 1S-3S transition}
\label{sec:2}

\subsection{Experimental setup}
\label{sec:21}

\begin{figure}
\resizebox{1.0\columnwidth}{!}{%
  \includegraphics{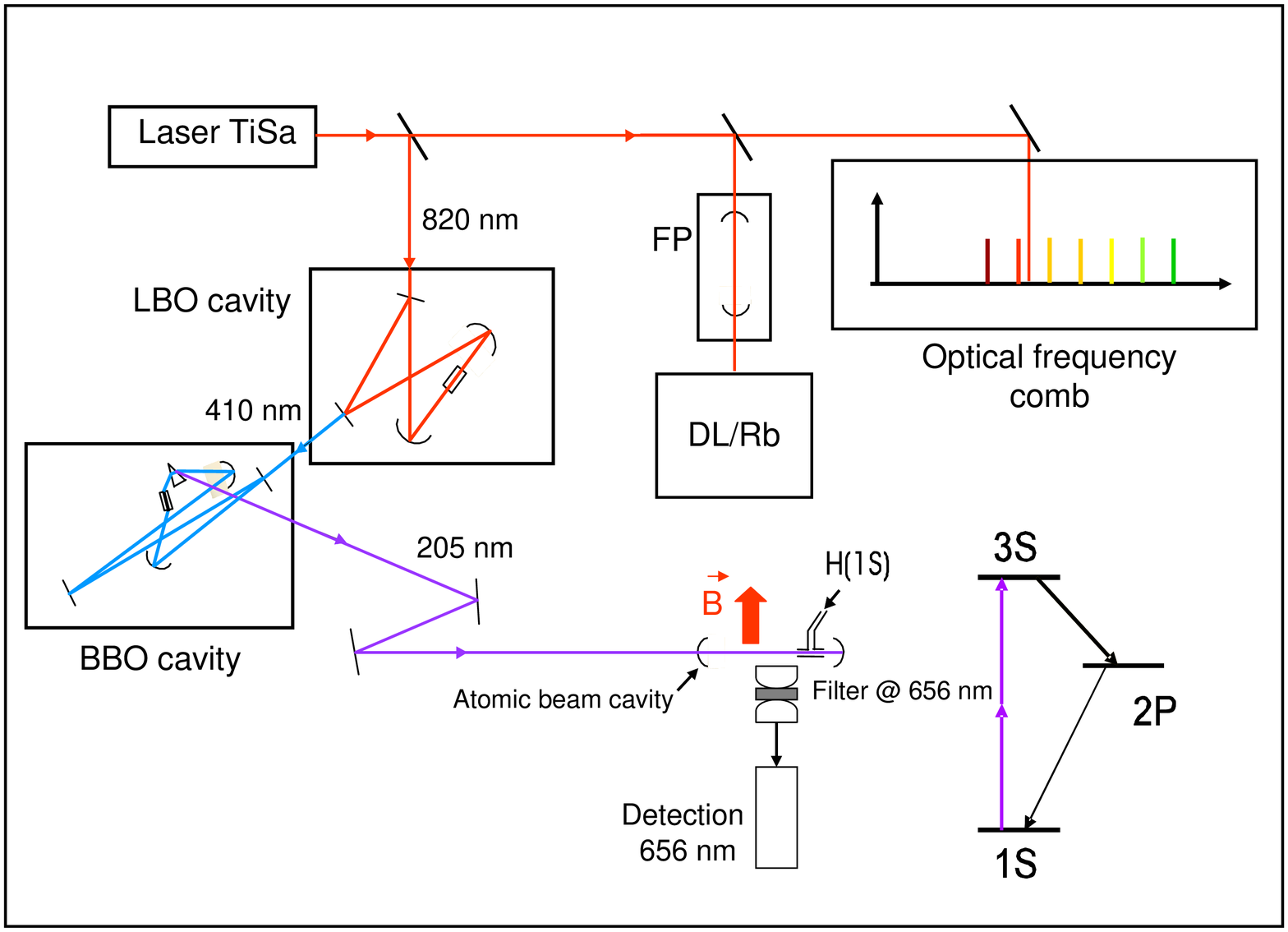}
}
\caption{Experimental setup for the excitation of the 1S-3S transition and the optical frequency measurement (Laser TiSa:  titanium sapphire laser, FP: Fabry-Perot cavity, DL/Rb: laser diode stabilized to a two-photon transition of Rubidium).}
\label{fig:1}       
\end{figure}

The experimental setup (see Figure \ref{fig:1}) has been described elsewhere \cite{{EPJD00},{Bourzeix},{Hagel}}. The excitation wavelength of the 1S-3S two-photon transition is in the UV range, at 205 nm. This radiation is produced by quadrupling in frequency a CW titanium sapphire laser at 820~nm, with two frequency doubling stages by a lithium triborate crystal (LBO) and a beta-barium borate crystal (BBO). These crystals are placed in two successive enhancement ring cavities (LBO cavity and BBO cavity respectively). The first frequency doubling is efficient and delivers about 800~mW at 410~nm from an incident power of 2~W at 820~nm \cite{LBO}. This first cavity is locked to the laser wavelength
without any length modulation thanks to the polarization method \cite{Hansch-Couillaud}. The second doubling step is by far more challenging \cite{BBO}. For the second harmonic generation at 410~nm, the only possible choice is a BBO crystal used at the limit of its phase-matching angle. This results in a low conversion efficiency, with a UV power below 1~mW. The second doubling step is operated in an $\mathrm{O}_2$ environment to slow the chemical reactions at the surface of the crystal. Furthermore, a photo-refractive effect takes place inside the crystal, which results in a reflected blue beam at 410~nm appearing from one end facet of the BBO crystal. A counter propagating wave at 410 nm develops in the ring BBO cavity. To reduce this effect, we have worked in a quasi-continuous regime where the UV intensity consists of $6~\mu \mathrm{s}$
pulses at a frequency of 30 kHz. This is done by overmodulating the length of the BBO cavity at the frequency $\nu _{0}$ of 15~kHz, so as to be resonant only twice per modulation period for 6 $\mu$s. First-order Doppler effect due to the motion of the cavity mirror results in frequency shifts, which induce a splitting of the observed 1S-3S line. This effect is described in the references \cite{{Hagel},{DeuxBosses}}.

\begin{figure}
\resizebox{0.9\columnwidth}{!}{%
  \includegraphics{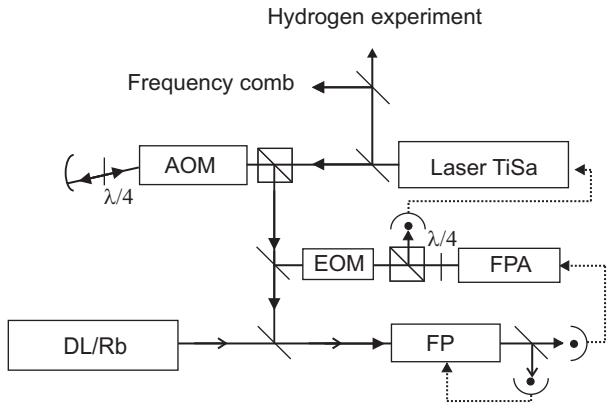}
}
\caption{Experimental setup for the frequency stabilization of the titanium sapphire laser. See the explanations in the text (TiSa: titanium sapphire laser, AOM: acousto-optic modulator, EOM: electro-optic modulator, FP and FPA:  Fabry-Perot cavities, DL/Rb: diode laser stabilized on a two-photon transition of Rb).}
\label{fig:2}       
\end{figure}

The frequency stabilization of the titanium sapphire laser is described in reference \cite{EPJD00}. The setup is shown in Figure \ref{fig:2}. The short term and long term stabilities are assured by two Fabry-Perot cavities, labelled FPA (auxiliary Fabry-Perot) and FP respectively. The principle of this stabilization arrangement is to lock the titanium-sapphire laser to the FPA
cavity, the FPA cavity to the FP cavity and, finally, the FP cavity to a diode laser stabilized on a two-photon transition of rubidium. A secondary laser beam from the titanium-sapphire laser is split after a double pass through an
acousto-optic modulator (model 3200 from Crystal Technology at 200~MHz) and sent on the FPA and FP cavities. The FPA cavity (free spectral range 600~MHz and finesse of about 400) is placed in a robust brass vacuum box (wall thickness of 2~cm) and carefully isolated from external vibrations \cite{LBO}. To reduce the frequency jitter, the laser is locked,
in a first step, to the FPA~cavity with a FM sideband method \cite{Drever-Hall}. Thanks to this servo-loop, the frequency jitter is reduced from 500~kHz (free running laser) to about 2~kHz \cite{LBO}.

The long term stability is guaranteed by the FP cavity. This cavity is very stable. It consists of a 50~cm long zerodur spacer and two silver coated mirrors, one flat and one spherical (60~cm curvature radius). Its finesse is about 120 at 800~nm. A piezoelectric transducer moves the flat mirror thanks to a mechanical construction (made in fused silica) which avoid the rotation of the mirror (the principle is to deform a parallelogram)~\cite{These Nez}. This cavity
is also placed in a vacuum box with the same design than for the FPA cavity. To obtain long term stability, the FP cavity is stabilized on a standard laser, namely a laser diode at 778~nm stabilized to the 5S$_{1/2}$-5D$_{5/2}$ two-photon transition of rubidium (DL/Rb laser). This standard has been described previously  \cite{{TowardsRb},{MesRubidium97},{RbLPTF}}. With this setup, as the zerodur spacer is very stable, the servo-loop of the FP cavity on the DL/Rb laser always uses the same fringe of the FP cavity and the length of the FP cavity is exactly known. Consequently, to excite the $1\mathrm{S}-3\mathrm{S}$ transition, the titanium sapphire laser is always locked on a fringe of the FP cavity which is also exactly known (fringe number 1219477 of the FP cavity) . The advantage of this method is that the 1S-3S signal always appears around the same frequency of the acousto-optic modulator. The frequency characteristics of this system will be presented in section \ref{sec:23}. Finally, to scan the laser frequency, we sweep the frequency of the radiofrequency wave which drives the acousto-optic modulator.

The 1S-3S transition is excited in a thermal 1S atomic beam colinear with the UV laser beams. The hydrogen atoms are produced through a radio-frequency discharge from molecular hydrogen. In order to increase the UV intensity, the atomic beam is placed inside a linear build-up cavity (UV cavity) formed by two spherical mirrors (radius of curvature 25~cm, 49~cm apart). Inside the cavity, the UV beam is focused within a waist of 48~$\mu $m. Because of the average characteristics of the UV mirrors (reflection and transmission of the input mirror 89~\% and 8.5~\%, reflection of the end mirror 96~\%), the cavity finesse is about 40. The mirrors are mounted on piezoelectric translators and the cavity length is locked to the 205~nm wavelength. This locking is detailed in reference \cite{DeuxBosses}. For this servo-loop, the amplitude of the frequency modulation due to the modulation of the BBO cavity is too small with respect to the width of the UV cavity to obtain a good error signal. Therefore the length of the UV cavity needs to be modulated at the same frequency as that of the BBO cavity (15~kHz), but in quadrature. This way, the UV pulses issued from the BBO cavity are sent to the UV cavity when the length of the UV cavity is at an extremum. If the UV cavity length is minimum for one pulse, it will be maximum for the following one and so on: the UV pulses test successively the two sides of the Airy's peak of the UV cavity. For this servo loop, the modulation amplitude is a small fraction of the width
of an Airy's peak of the UV cavity (typically 20~\%). Finally, the UV intensity inside the cavity is monitored by a photodiode placed after the end mirror and a phase sensitive detection at 15 kHz compares the transmission of the UV cavity for two successive pulses to obtain the servo signal.

The two-photon transition is detected by monitoring the Balmer-$\alpha $ fluorescence due to the radiative decay 3S-2P. This fluorescence is collected with a spherical mirror and a $f/0.5$ aspheric lens system and selected with an interference filter at 656 nm. We have used two different detectors, a photomultiplier (Hamamatsu R943-02) or a CCD camera (Princeton Instruments Spec 100B). The following section presents the data acquisition in these two cases.

\subsection{Experimental spectra}
\label{sec:22}

\begin{figure}
\resizebox{1.0\columnwidth}{!}{%
  \includegraphics{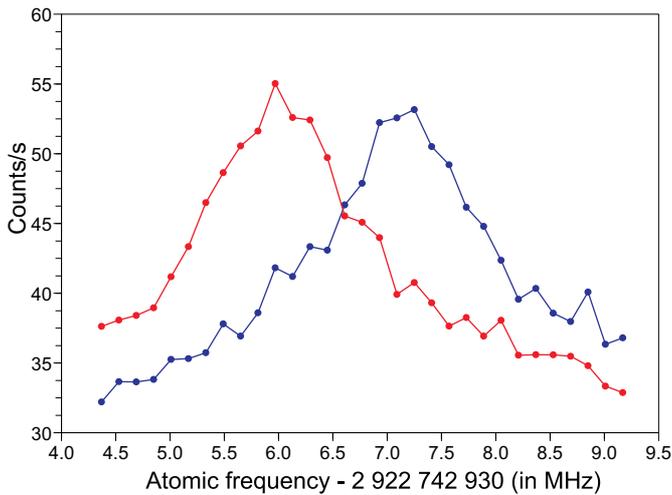}
}
\caption{Record of the 1S-3S two-photon transition in hydrogen with the photomultiplier (color online). This spectrum is the mean of 8 runs. For each point the acquisition time is 80 s. The two curves correspond to the signal recorded when the BBO cavity length increases (blue curve, blue shifted) or decreases (red curve, red shifted).}
\label{fig:3}       
\end{figure}

For the photomultiplier detection, the data acquisition takes benefit from the short time response of the photomultiplier. Each scan is divided in 31 frequency points. For each point, the photomultiplier signal is counted during 1~s and we make 10 scans of the line to achieve a 7~minutes run. Furthermore, acquisition electronics are designed to select only the time window during which the 205~nm UV light is resonant inside the atomic beam cavity (as a consequence of the 15 kHz over modulation of the BBO cavity). The signal of the photomultiplier is sent, after an amplifier discriminator and an electronic switch, to a multiplexer. To reduce the noise due to the dark current of the photomultiplier, the electronic switch transmits the signal only when the instantaneous UV intensity is above a reference level. Then, a multiplexer dispatches the signal to two counters in phase with the modulation at 15 kHz: when the BBO cavity length increases, the signal is sent to counter~1, and, when it decreases, to counter~2. This way, we can observe the splitting of the $1\mathrm{S}-3\mathrm{S}$ line due to the Doppler shift induced by the motion of the mirror (see section \ref{sec:21}). As the signal-to-noise ratio is small, we take the mean of several runs to obtain an observable signal. Figure~\ref{fig:3} shows the atomic signal. The two curves, corresponding to the two counters are separated by about $1.1~ \mathrm{MHz}$ in terms of atomic frequency. Consequently we will have to take into account this effect in the line shape analysis.

\begin{figure}
\resizebox{1.0\columnwidth}{!}{%
  \includegraphics{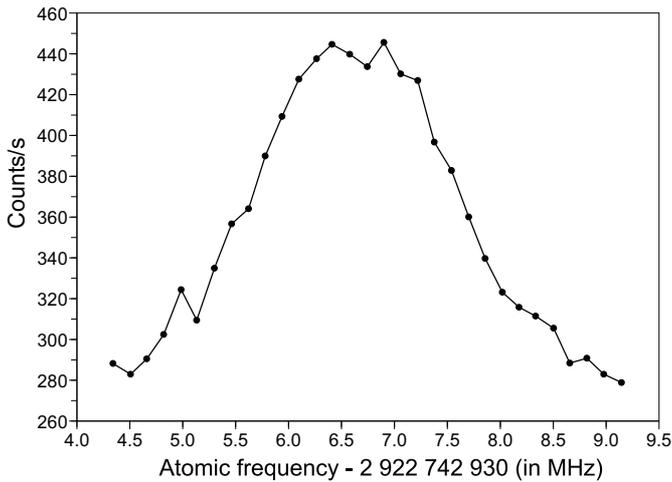}
}
\caption{Record of the 1S-3S two-photon transition in hydrogen with a CCD camera. The total acquisition time is about 20 minutes. In this case the signals obtained when the BBO cavity length increases or decreases are not separated.}
\label{fig:4}       
\end{figure}

In comparison with the photomultiplier, the CCD camera has two advantages: the quantum efficiency is higher (about 90 \% for a back-illuminated camera when it is 15 \% for the photomultiplier) and it is possible to obtain an image of the fluorescence of the atoms in the laser beam. The drawback is the longer time response of the CCD, of the order of one millisecond, which does not allow us to perform the electronic time selection described above. Moreover a readout noise, independent of exposure time, is also superimposed on each pixel (or group of pixels) every time the chip is being read. This leads to the use of long exposure times, of the order of 1 minute, to gather as many signal photons as possible before reading the chip. Nevertheless, this exposure time should remain small compared to the characteristic drifting time of the titanium sapphire laser and of the frequency doubling stages. Finally the exposure time of each frequency point of the scan is 37 s and the acquisition of the 31 frequency points lasts about twenty minutes. Then each image is carefully analyzed to reduce the parasitic signal due to the UV and to take into account the variation of the UV intensity during the scan (see the references \cite{{Arnoult},{TheseArnoult}} for the detail of this analysis). An example of $1\mathrm{S}-3\mathrm{S}$ spectrum obtained with the CCD camera is shown in figure \ref{fig:4}. For this record the UV intensity was about 70 \% the one of the record shown in figure \ref{fig:3}. As the two-photon excitation is a quadratic process, this corresponds to a reduction by a factor of about 2 of the excitation probability. Nevertheless the CCD camera signal is about five times larger than the photomultiplier signal. This corresponds to the gain in the quantum efficiency and to a better collection of the atomic fluorescence. For this reason, the main results presented in this article have been obtained by using the CCD camera.

\subsection{Optical frequency measurement}
\label{sec:23}

\begin{figure}
\resizebox{1.0\columnwidth}{!}{%
  \includegraphics{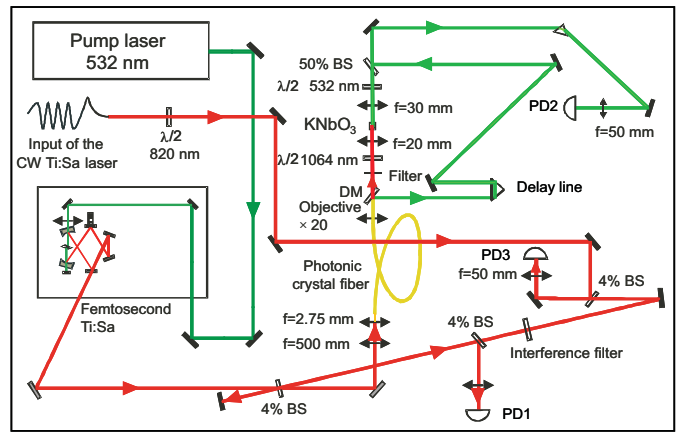}
}
\caption{Schematic diagram of the optical setup for the optical frequency measurements (Ti:Sa: titanium-sapphire laser, BS: beam splitter, DM: dichroic mirror, PD: photodiode). The repetition rate is detected with PD1, the offset $f_0$ with PD2 and the beat note between the CW Ti:Sa laser and the frequency comb with PD3 (color online).}
\label{fig:5}       
\end{figure}

The optical frequency measurements are made with an optical frequency synthesizer, following the design introduced by Hall and H\"{a}nsch \cite{{Femto1},{Femto2}}. Figure \ref{fig:5} shows the experimental arrangement. We use a second titanium sapphire laser, a six chirped mirrors mode-locked femtosecond laser with a repetition rate $f_{rep}$ of about 900~MHz (laser GigaJet 20 from Menlo Systems GmbH) with a 5~W pump laser (laser Verdi V5 from Coherent). The output of the titanium sapphire laser forms a frequency comb. The frequency $f_N$ of each line of the comb is controlled by the frequency rate $f_{rep}$ and the global shift $f_0$ of the frequency comb with respect to the zero frequency: $f_N=Nf_{rep}+f_0$. The offset frequency $f_0$ is determined thanks to the self-referencing technique. The spectrum of the femtosecond titanium sapphire laser is broadened inside a microstructure photonic crystal fiber (from CrystalFiber) as to span over more than an octave. Then the infrared part of the spectrum is frequency doubled in a KNbO$_3$ nonlinear crystal to obtain green radiation which is recombined, after a time delay, with the green part of the spectrum generated by the photonic crystal fiber. The result is a beat note detected by the photodiode PD2 (see figure \ref{fig:5}) at the offset frequency $f_0$ or at the frequency $f_{rep}-f_0$. As the output spectrum of the comb before the fiber (ranging from 780 to 820~nm) includes the wavelength of the two-photon excitation, the CW titanium-sapphire laser is mixed with the frequency comb before the photonic crystal fiber. The advantage is that the resulting beat note (frequency $f_1$) is very stable with a good signal-to-noise ratio (40 dB in a 300 kHz bandwidth). This frequency $f_1$ corresponds to the frequency difference between the titanium sapphire laser frequency and the closest frequency of the comb, either lower (case (a)) or higher (case (b)). The spacing between the lines of the comb is fixed by phase-locking the repetition rate to a reference signal related to a Cs clock. Thanks to an optical link between our laboratory and LNE ({\it Laboratoire National d'Essais})-SYRTE {\it (Syst\`{e}me de R\'{e}f\'{e}rence Temps Espace}) at the {\it Observatoire de Paris} \cite{Fibre}, we receive a radio-frequency signal at 100~MHz locked to a primary frequency standard. Then a radio frequency chain generates a signal at 11~GHz which is mixed with the twelfth harmonic of the repetition rate \cite{FemtoSyrte}. The phase error signal is then amplified and fed back to a piezoelectric translator controlling the length of the femtosecond ring cavity. Finally the frequency of the titanium-sapphire laser is given by:
\begin{equation}
f_{\mathrm{Ti:Sa}}=N \times f_{rep}+ f_0 \pm f_1  \label{Eq1}
\end{equation}
with a $+$ sign in case (a) and $-$ sign in case (b). Then, by mixing both beat notes electronically and filtering, one can record the suitable combination of $f_0$ and $f_1$ with a tracking oscillator and a frequency counter.

\begin{figure}
\resizebox{1.0\columnwidth}{!}{%
  \includegraphics{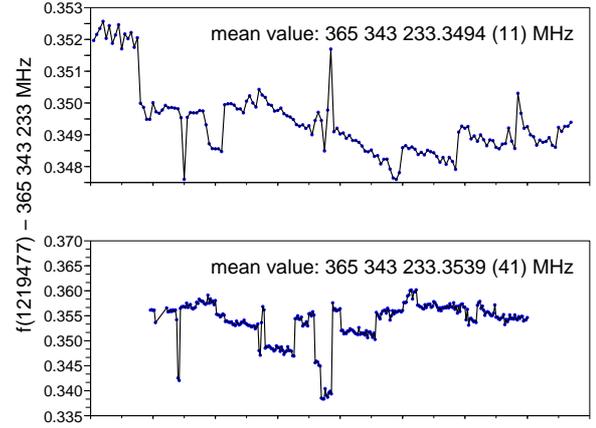}
}
\caption{Measurements of the frequency $f(1219477)$ of the peak numbered 1219477 of the FP cavity, in 2005 (upper part) and in 2009 (lower part). The horizontal axis corresponds to the successive number of the runs (about 200 runs during 15 days in 2005 and 300 runs during 30 days in 2009). Each point corresponds to a 20 minute data acquisition run in 2005, and 7 minute in 2009. During this period, the drift of the cavity is of a few kHz.}
\label{fig:6}       
\end{figure}

Thanks to the optical link with the Observatoire de Paris, the DL/Rb standard laser has been simultaneously measured during 300 s in our laboratory and in the LNE-SYRTE laboratory. The result of this test is a frequency difference of $2~(26)~\mathrm{Hz}$, the uncertainty corresponds to one standard deviation of the mean. The absolute frequency of the DL/Rb standard has also been measured several times. The results obtained after extrapolation to zero laser intensity (in order to eliminate the light shift of the two-photon rubidium transition) are $385~285~142~376.3~(1.0)$ $\mathrm{kHz}$ in May 2004, $385~285~142~376.4~(1.0)~\mathrm{kHz}$ in June 2005 and $385~285~142~378.8~(1.0)~\mathrm{kHz}$ in February 2009. These values are close to the first measurement of the DL/Rb standard en 1996 which was $385~285~142~376.7~(1.0)~\mathrm{kHz}$ \cite{{MesRubidium97},{MesCO2-3}}.

During the recordings of the 1S-3S spectra, the frequency of the titanium-sapphire laser was continuously measured. Figure \ref{fig:6} shows the optical frequency of the peak of the FP cavity numbered 1219477 which is used to lock the titanium-sapphire laser, measured in 2005 and in 2009. This figure illustrates the repeatability of the setup used to stabilized the titanium-sapphire laser (see figure \ref{fig:2}). The frequency jumps are due to some degradations of the servo-loops of the lasers on the FP cavity. Usually it is due to a bad alignment of the lasers with respect to the cavity or to a diminution of the electronic gain. The shift of about 5~kHz between the 2005 and 2009 measurements can be explained by the frequency of the DL/Rb laser or by an ageing of the silver coated mirrors of the FP cavity.
\section{Determination of the velocity distribution}
\label{sec:3}

\begin{figure}
\resizebox{1.1\columnwidth}{!}{%
  \includegraphics{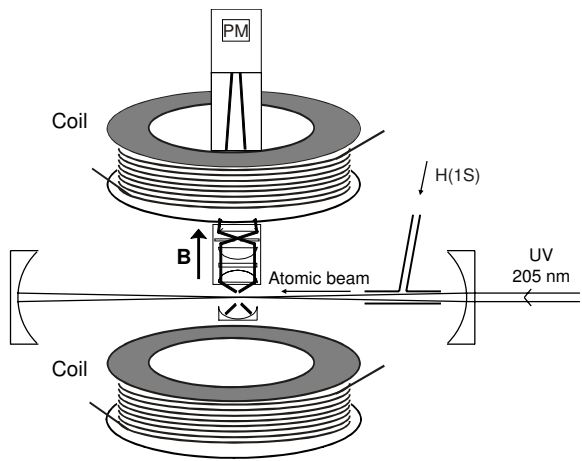}
}
\caption{Experimental arrangement for the determination of the second-order Doppler shift. The atomic beam is horizontal. The UV light is focused in front of the detection system using a photomultiplier (PM). The two coils produce a vertical magnetic field ${\bf B}$ perpendicular to the direction of the atomic beam.}
\label{fig:7}       
\end{figure}

In a thermal hydrogen beam at room temperature, the typical atomic velocity is 3 km/s. For the 1S-3S transition, this velocity induces a second-order Doppler shift $-\nu(1{\rm S}-3{\rm S}) \times v^2/2c^2$ which is about 146 kHz ($\nu(1{\rm S}-3{\rm S})$ is the atomic frequency). To measure this effect, we use a method proposed in reference \cite{Doppler} which is convenient when the 1S-$n$S (or 2S-$n$S) two-photon hydrogen transitions are produced in an atomic beam. The principle is to
apply a transverse magnetic field ${\bf B}$ with respect to the direction of the atomic beam (see figure \ref{fig:7}). This magnetic field produces a motional electric field ${\bf E} = {\bf v} \times {\bf B}$ which induces a quadratic Stark shift proportional, as the second-order Doppler effect, to ${v}^{2}$. From the observation of this effect, we can deduce the second order Doppler shift.

\begin{figure}
\resizebox{1.0\columnwidth}{!}{%
  \includegraphics{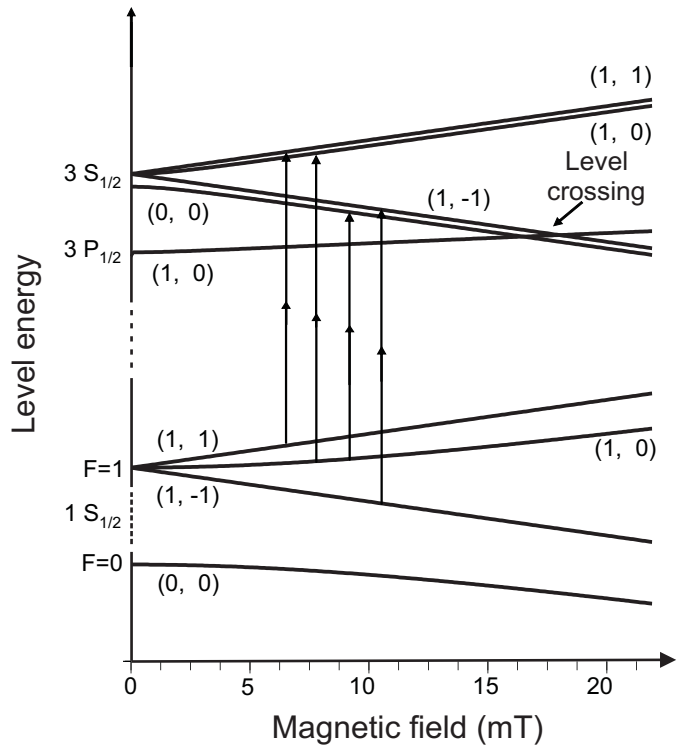}
}
\caption{Zeeman diagram of the 1S-3S transition. The levels are labelled by the quantum numbers ($F,m_{F}$). In the intermediate regime the 1S$_{1/2}$-3S$_{1/2}(F=1)$ line is split in four components following the selection rule $\Delta m_{F}=0$.
The motional Stark effect is important, for the 1S$_{1/2}$-3S$_{1/2}$($F=1,m_{F}=-1$) component, around the crossing between the 3S$_{1/2}$ ($F=1,m_{F}=-1$) and 3P$_{1/2}$($F=1,m_{F}=0$) levels. For high magnetic field, the magnetic quantum numbers of these two levels are respectively ($m_J=-1/2$, $m_{I}=-1/2$) and ($m_J=1/2$, $m_{I}=-1/2$).}
\label{fig:8}       
\end{figure}

The first effect of the magnetic field is a Zeeman splitting (see figure \ref{fig:8}). In low magnetic field, because of the selection rules of a two-photon transition ($\Delta F=0$ and $\Delta m_{F}=0$ \cite{JPhys73}), the line between the hyperfine levels 1S$_{1/2}(F=1$) and 3S$_{1/2}(F=1$) is split in three lines. For a higher magnetic field, the line between the 1S$_{1/2}(F=1, m_{F}=0$) and  3S$_{1/2}(F=0, m_{F}=0$) Zeeman sub-levels becomes allowed, because the hyperfine structure of the 3S level is in the Paschen-Back regime when it is not the case for the 1S level. Practically, the shift of the $m_{F}=\pm 1$ components is very small, because, in first approximation, the Land\'{e} factors are the same for the 1S and 3S levels. Following a relativistic calculation, the Land\'{e} factors $g_{J}$ of the 1S and 3S levels are respectively 2.002284 and 2.002315 \cite{Paul}. This difference induces, for a field of 20 mT, a residual Zeeman shift of about $\pm 4.3$~kHz. For this field the diamagnetic shift is about 809 Hz.  On the other hand, the $F=1, m_{F}=0$ component is shifted, because the Zeeman effect is not the same for the 1S$_{1/2}(F=1,m_{F}=0$) and 3S$_{1/2}(F=1,m_{F}=0$ ) sub-levels. For 20~mT, this shift is about 200~MHz.

The second effect of this magnetic field is a motional electric field ${\bf E=v\times B}$. For a velocity of 3 km/s and a magnetic field of 20 mT, this electric field is 0.6 V/cm. It induces a coupling of the 3S$_{1/2}$ level with the 3P$_{1/2}$ and 3P$_{3/2}$ levels. For the 3S$_{1/2}$ level, in the limit of high magnetic field, the electronic and nuclear spins are decoupled and the good quantum numbers are ($J$, $I$, $m_J$, $m_I$). For example, the 3S$_{1/2}$($F=1,m_{F}=-1$) level corresponds to the quantum numbers $m_J=-1/2$ and $m_I=-1/2$. It is mainly coupled to the 3P$_{1/2}$($m_J=1/2$, $m_{I}=-1/2$), 3P$_{3/2}$($m_J=1/2$, $m_{I}=-1/2$) and 3P$_{3/2}$($m_J=-3/2$, $m_{I}=-1/2$) sub-levels following the selection rules $\Delta m_J= \pm 1$ and $\Delta m_I=0$. This kind of motional Stark mixing is well known and, for instance, was used by Lamb and Retherford to polarize a metastable atomic beam \cite{Lamb}. We consider the quadratic Stark effect due to this motional electric field. This shift, proportional to ${\bf v}^{2}$, is very small, but observable in the case of the $m_{F}=\pm 1$ line components, because the Zeeman shift of these components is negligible. In particular, this effect is important around the level crossing between the 3S$_{1/2}$($F=1,m_{F}=-1$) and 3P$_{1/2}$($F=1,m_{F}=0$) Zeeman sub-levels, which appears for a magnetic field of about 18 mT (see figure \ref{fig:8}).

The complete calculation of the line shape will be described in the section \ref{sec:4}. We present here a simple picture. The initial state $g$ (the levels 1S$_{1/2}$($F=1,m_{F}=\pm 1$)) is coupled by two-photon excitation
to an excited state $e$ (here the levels 3S$_{1/2}$($F=1,m_{F}=\pm 1$)). This state $e$ is mixed by the Stark hamiltonian $V_{ef}$ with several $f$ states (the Zeeman sub-levels of the 3P$_{1/2}$ and 3P$_{3/2}$ states). In a simple approach, the perturbed energy $E^\prime_e$ and width $\Gamma^\prime_e$ of the considered $e$ level are deduced from the unperturbed energy and width $E_e$ and $\Gamma_e$, using the perturbation theory in lowest order:
\begin{equation}
E^\prime_e-i\frac{\hbar}{2} \Gamma^\prime_e= E_e-i\frac{\hbar}{2} \Gamma_e+\sum_{f}\frac{\left| V_{ef}\right| ^{2}}{( E_e-E_f)-i\frac{\hbar}{2}(\Gamma_e-\Gamma_f)} \label{Eq2}
\end{equation}
Then the Stark shift $\delta_S $ of the level $e$ is given by:
\begin{equation}
\delta_S=\mathop{\rm Re}(E^\prime_e-E_e)=\sum_{f}\frac{( E_e-E_f)\left| V_{ef}\right|^{2}} {(E_e-E_f)^2+(\frac{\hbar}{2})^2(\Gamma_e-\Gamma_f)^2} \label{Eq3}
\end{equation}
In these equations, the energies $E_e$ and $E_f$ take into account the shifts of the $e$ and $f$ levels due to the Zeeman effect: then $E_e-E_f$ is a function of the magnetic field and the Stark shift appears as a sum of dispersion curves corresponding to the different level crossings between the 3S$_{1/2}$ and 3P$_J$ levels. There is no divergence at the level crossing when $E_e=E_f$ because of the difference between the natural widths of the 3S$_{1/2}$ and 3P$_{1/2}$ levels (respectively 1 MHz and 30.6 MHz). As the effect of the electric field is negligible for the 1S$_{1/2}$ level, the total shift $\delta $ of the two-photon line is:
\begin{equation}
\delta = \delta_Z+\delta_S-\nu(1{\rm S}-3{\rm S}) \times \frac{v^2}{2c^2} \label{Eq4}
\end{equation}
where $\delta_Z $ is the shift due to the difference of the Zeeman effect for the $g$ and $e$ levels and the third term the second Doppler effect for an atom of velocity $v$.

\begin{figure}
\resizebox{1.0\columnwidth}{!}{%
  \includegraphics{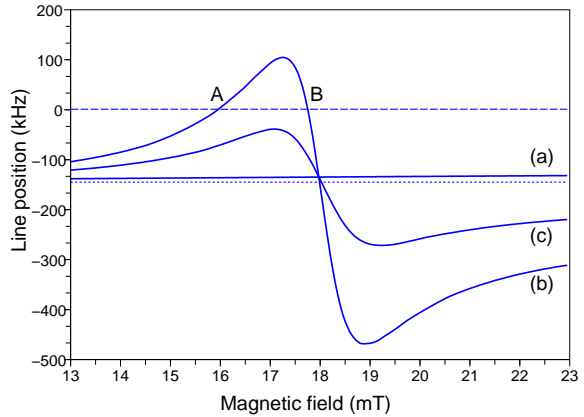}
}
\caption{Shift of the 1S-3S line due to the motional electric field (calculation for an atom at a velocity of 3~km/s). Zero: line position for an atom at rest; small dashed line: shift of the line due to the second-order Doppler effect. Curves (a) and (b): line position of the 1S$_{1/2}$-3S$_{1/2}$($F=1,m_{F}=+1$) and 1S$_{1/2}$-3S$_{1/2}$($F=1,m_{F}=-1$) components. In A and B, the second-order Doppler effect of the 1S$_{1/2}$-3S$_{1/2}$($F=1,m_{F}=-1$) component is exactly compensated by the motional Stark effect. Curve (c): global motional Stark effect for the two components
1S$_{1/2}$-3S$_{1/2}$($F=1,m_{F}=\pm 1$).}
\label{fig:9}       
\end{figure}

Figure \ref{fig:9} shows, for an atom at 3~km/s, the shifts of the 1S$_{1/2}-$3S$_{1/2}(F=1,m_{F}=\pm 1$) transitions around the level crossing at 18 mT. The large dashed line represents the line position without any shifts for the velocity $v=0$ and the small dashed line the line position shifted by the second-order Doppler effect for $v=3~\mathrm{km/s}$ and $B=0$. This shift is about $-146~\mathrm{kHz}$. With respect to this reference, the shift of the 1S$_{1/2}$-3S$_{1/2}$($F=1,m_{F}=+1$) line (curve (a)) is small (about 10~kHz), because, in this range of magnetic field, the 3S$_{1/2}$($F=1,m_{F}=+1$) level is far from the 3P$_{1/2}$ and 3P$_{3/2}$ levels. On the other hand, the shift of the 1S$_{1/2}$-3S$_{1/2}$($F=1,m_{F}=-1$) line (curve (b)) is
important around the level crossing. It has a dispersion shape and is larger than the second-order Doppler effect. Moreover, there are two values of the magnetic field (points A and B) where the second-order Doppler effect is
exactly compensated by this motional Stark effect. For these values, the position of the line is independent of the atomic velocity, because the second-order Doppler effect and the motional Stark effect both vary as $v^2$.
Nevertheless, this effect cannot be used directly, because the 1S$_{1/2}$-3S$_{1/2}$($F=1,m_{F}=\pm 1$) lines are not separated (the motional Stark effect is small with respect to the 3S$_{1/2}$ natural width (about 1~MHz). Curve (c) shows the barycenter of these two lines. Though reduced by a factor two, the dispersion amplitude is about 230~kHz, and the compensation of the second-order Doppler effect is still 73\% for a magnetic field of 17~mT. We have observed this effect to deduce the second-order Doppler effect.

\section{Line shape}
\label{sec:4}

\subsection{Theoretical background }
\label{sec:41}

The aim of this calculation is to take into account simultaneously the natural width and the Zeeman and Stark effects. We follow the procedure described in reference \cite{EPJD00}. The evolution of the density operator $\rho $ is:
\begin{equation}
\frac{d\rho }{dt}=\frac{1}{i\hbar }\left[ \left( H_{0}+V_{L}+V_{S}\right)
,\rho \right] +\Gamma \rho  \label{Schrodinger}
\end{equation}
where $V_{S}$ is the Stark Hamiltonian and the operators $V_{L}$ and $\Gamma $ describe the two-photon excitation and the spontaneous emission. We use the notations of section \ref{sec:3}: the states $g$, $e$ and $f$ are respectively the  1S$_{1/2}$, 3S$_{1/2}$ and 3P$_J$ levels. We make the rotating wave approximation and we introduce the two-photon Rabi frequency $\Omega _{e}$:
\[
\left\langle e\right| V_{L}\left| g\right\rangle =\frac{\hbar \Omega _{e}}{2}\exp(-2i\omega t)
\]
\begin{equation}
\Omega _{e}=\frac{8\pi a_{0}^{2}\left| \langle e\left| Q_{tp}\right|
g\rangle \right| I}{mc^{2}\alpha }  \label{Rabi2ph}
\end{equation}
where $\omega /2\pi$ is the laser frequency, $I$ the power density of the light, $a_{0}$ the Bohr radius, $\alpha $ the fine structure constant and $m$ the electron mass. For a polarization along the z-axis, the two-photon operator $Q_{tp}$ is given in atomic units ($\hbar =\alpha c=m=1$) by:
\begin{equation}
Q_{tp}=\sum_{r} \frac{z\left| r\right\rangle \left\langle r\right| z}{\omega -\omega _{rg}}
\label{Qtwo-photon}
\end{equation}
where the sum is made on all the atomic states $r$ and $\omega _{rg}/2\pi$ is the atomic frequency difference between the $g$ and $r$ states. Because of the selection rules for the two-photon excitation, the initial state $g$ is coupled to a single excited state $e$.

If we assume that $\Omega _{e}\ll \Gamma _{e}$, we can neglect in a first step the populations and coherences $\rho _{ee^{\prime }}$, $\rho _{ff^{\prime }}$ or $\rho _{ef}$ of the upper levels. In the rotating frame, we replace the density operator by an operator $\sigma $ with $\sigma _{gg}=\rho _{gg}$, $\sigma _{eg}=\rho _{eg}\exp (2i\omega t)$ and $\sigma _{ge}=\rho _{ge}\exp
(-2i\omega t)$ and we introduce the frequency detunings $\Delta _{e}=2\omega-(\omega _{e}-\omega _{g})$ and $\Delta _{f}=2\omega -(\omega _{f}-\omega _{g})$ ($\hbar\omega_i$ is the energy of the level $i$). In this way, we obtain from equation (\ref{Schrodinger}) a set of equations:
\begin{equation}
\frac{d\sigma _{gg}}{dt}=-\frac{i}{2} \Omega _{e}\left( \sigma_{eg}-\sigma _{ge}\right)
\label{Population}
\end{equation}
\begin{equation}
\frac{d\sigma _{eg}}{dt}=\left( i\Delta _{e}-\frac{\Gamma _{e}}{2}\right)\sigma _{eg}
-i\frac{\Omega _{e}}{2}\sigma _{gg}-\frac{i}{\hbar }\sum\limits_{f}V_{ef}\sigma _{fg}
\label{Coherence1}
\end{equation}
\begin{equation}
\frac{d\sigma _{fg}}{dt}=\left( i\Delta _{f}-\frac{\Gamma _{f}}{2}\right)\sigma _{fg}
-\frac{i}{\hbar }V_{fe}\sigma _{eg}
\label{Coherence2}
\end{equation}
where $V_{fe}=\left\langle f\right|V_{S}\left| e\right\rangle $. Then we assume that the optical coherences $\sigma _{eg}$ and $\sigma _{fg}$ follow adiabatically the population $\sigma_{gg}$, {\it i.e.} that:
\[
\frac{d\sigma _{eg}}{dt}=0\qquad {\rm and\qquad }\frac{d\sigma _{fg}}{dt}=0
\]
With these hypotheses, the equations (\ref{Coherence1}, \ref{Coherence2})
give:
\begin{equation}
\left( i\Delta _{e}-\frac{\Gamma _{e}}{2}\right) \sigma_{eg}
+\sum\limits_{f}\frac{V_{ef}V_{fe}}{\hbar^{2}\left( i\Delta_{f}-\frac{\Gamma _{f}}{2}\right) }\sigma _{eg}
=i\frac{\Omega _{e}}{2}\sigma _{gg}
\label{Coherence3}
\end{equation}
This equation gives the coherence $\sigma_{eg}$ as a function of the population $\sigma_{gg}$. It can be written more simply:
\begin{equation}
\sigma_{eg}=\frac{1}{A}i\frac{\Omega _{e}}{2}\sigma _{gg}
\label{Coherence4}
\end{equation}

where:
\begin{equation}
A=\left( i\Delta _{e}-\frac{\Gamma _{e}}{2}\right)
+\sum\limits_{f}\frac{V_{ef}V_{fe}}{\hbar^{2}\left( i\Delta_{f}-\frac{\Gamma _{f}}{2}\right) }
\label{Coherence5}
\end{equation}
Combining with equation (\ref{Population}), one obtains the evolution of the population $\sigma_{gg}$ and the probability of the two-photon excitation.

To obtain the correct line shape, we have to calculate the 656 nm fluorescence from the $n=3$ levels towards the $n=2$ levels. This fluorescence is due to the cascades 3S$_{1/2}$ $\rightarrow$ 2P$_{J}$ and 3P$_{J}$ $\rightarrow$ 2S$_{1/2}$. Because of the Stark mixing, the 3P$_{J}$ states are also populated. Then, to obtain the line shape, it is required to calculate the populations $\sigma _{ee}$, $\sigma _{ff}$ and $\sigma _{ff^{\prime }}$. They are deduced from equation \ref{Schrodinger}:
\begin{eqnarray}
\frac{d\sigma _{ee}}{dt}&=&-\Gamma _{e}\sigma _{ee}+\frac{i}{2} \Omega _{e}\left( \sigma_{eg}-\sigma _{ge}\right) \nonumber \\
&&-\frac{i}{\hbar}\sum\limits_{f}\left(V_{ef} \sigma_{fe}-\sigma _{ef}V_{fe}\right)
\label{Population1}
\end{eqnarray}
\begin{eqnarray}
\frac{d\sigma _{fe}}{dt}&=&-\frac{\Gamma _{e}+\Gamma _{f}}{2}\sigma _{fe}
-i\left(\omega_f-\omega_e\right)\sigma _{fe}  \nonumber \\
&&+\frac{i}{2} \Omega _{e} \sigma_{fg}
-\frac{i}{\hbar}V_{fe}\left(\sigma_{ee}-\sigma_{ff}\right)
\label{Population2}
\end{eqnarray}
\begin{equation}
\frac{d\sigma _{ff}}{dt}=-\Gamma _{f}\sigma _{ff}-\frac{i}{\hbar}\left(V_{fe} \sigma_{ef}-\sigma _{fe}V_{ef} \right)
\label{Population3}
\end{equation}
In our experimental conditions, the two-photon excitation probability is very small, typically $10^{-3}$/s for an atom at the center of a 1 mW UV beam. As this probability is very small with respect to the natural width of the 3S level (1 MHz), we are in a stationary regime and $d\sigma _{ii^{\prime}}/dt=0$ (with $i$ and $i^{\prime}$ are the levels $e$ or $f$). Then it is possible to calculate the populations $\sigma_{ii^{\prime}}$ from the set of equations (\ref{Population1}$-$\ref{Population3}).

First we consider the simple case where the $e$ level is coupled with only one $f$ level by the Stark Hamiltonian. From equations (\ref{Coherence2}), (\ref{Coherence4}) and (\ref{Population2}), we deduce:
\begin{equation}
\sigma _{fe}=\frac{\frac{1}{\hbar}V_{fe}\left(\sigma_{ee}-\sigma_{ff}\right)-i\frac{\Omega_e^2}{4\hbar}\frac{1}{A}
\frac{V_{fe}}{\Delta_f+i\Gamma_f/2}\sigma_{gg}}
{\omega_e-\omega_f+i\frac{\Gamma_e+\Gamma_f}{2}}
\label{population4}
\end{equation}
The populations $\sigma_{ee}$ and $\sigma_{ff}$ are obtained from the equations (\ref{Population1}) and (\ref{Population3}):
\begin{eqnarray}
\Gamma _{e}\sigma _{ee}&=&2\mathop{Re}\left(i\frac{\Omega _{e}}{2}\sigma_{eg}\right)
-\frac{2}{\hbar}\mathop{Re}\left(iV_{ef}\sigma_{fe}\right)\nonumber \\
\Gamma _{f}\sigma _{ff}&=&\frac{2}{\hbar}\mathop{Re}\left(iV_{ef}\sigma_{fe}\right)
\label{Population5}
\end{eqnarray}
Using the expressions of $\sigma_{eg}$ (equation (\ref{Coherence4})) and $\sigma_{fe}$ (equation (\ref{population4})), we can express $\sigma_{ee}$ and $\sigma_{ff}$ as a  function of $\sigma_{gg}$:
\begin{equation}
\left(\begin{array}{c}\Gamma _{e}\sigma _{ee} \\ \Gamma _{f}\sigma _{ff}\end{array}\right)=
B\left(\begin{array}{cc}1&-1 \\ -1&1\end{array}\right)\left(\begin{array}{c}\sigma_{ee}\\\sigma_{ff}\end{array}\right)
+\left(\begin{array}{c}C_1 \\ C_2 \end{array}\right)\sigma_{gg}
\label{Population6}
\end{equation}
where the coefficients $B$, $C_1$ and $C_2$ are:
\begin{eqnarray}
B&=&-\frac{\frac{1}{\hbar^2}\vert V_{ef}\vert^2\left(\Gamma_e+\Gamma_f\right)}
{\left(\omega_e-\omega_f\right)^2+\left(\frac{\Gamma_e+\Gamma_f}{2}\right)^2}  \nonumber \\
C_1&=&-\frac{\Omega_e^2}{2}\mathop{Re}\left(\frac{1}{A}\left[1+\frac{\vert V_{ef}\vert^2/\hbar^2}
{\left(\Delta_f+i\frac{\Gamma_f}{2}\right)\left(\omega_e-\omega_f+i\frac{\Gamma_e+\Gamma_f}{2}\right)}\right]\right)
\nonumber \\
C_2&=&\frac{\Omega_e^2}{2}\mathop{Re}\left(\frac{1}{A}\frac{\vert V_{ef}\vert^2/\hbar^2}
{\left(\Delta_f+i\frac{\Gamma_f}{2}\right)\left(\omega_e-\omega_f+i\frac{\Gamma_e+\Gamma_f}{2}\right)}\right)
\label{coeffB}
\end{eqnarray}
Finally, the populations $\sigma_{ee}$ and $\sigma_{ff}$ are:
\begin{equation}
\left(\begin{array}{c}\sigma _{ee} \\ \sigma _{ff}\end{array}\right)=
\left(\begin{array}{cc}\Gamma_e-B&B \\B&\Gamma_f-B\end{array}\right)^{-1}
\left(\begin{array}{c}C_1 \\ C_2 \end{array}\right)\sigma_{gg}
\label{Population7}
\end{equation}
The calculations are similar in the case where there are several $f$ levels mixed by the electric field with the $e$ level. They are described in detail in the reference \cite{These Hagel}.

\subsection{Theoretical line shape}
\label{sec:42}

\begin{figure}
\resizebox{1.0\columnwidth}{!}{%
  \includegraphics{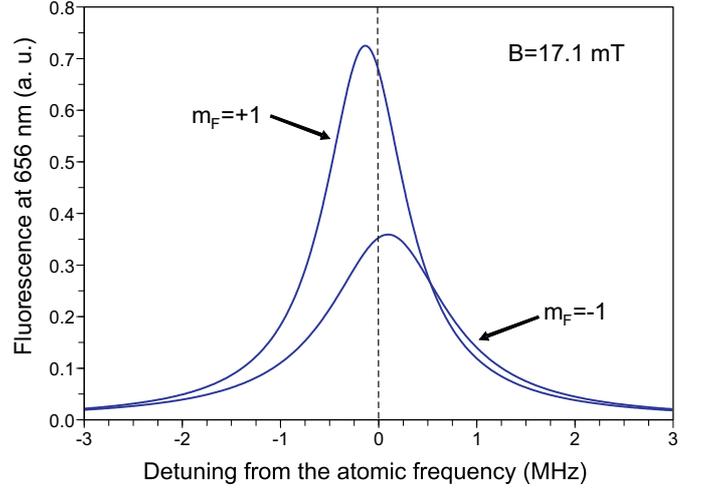}
}
\caption{Line shape of the $m_F=+1$ and $m_F=-1$ Zeeman components of the line for an atom at a velocity of 3~km/s in a magnetic field of 17.1 mT. The dashed line is the line position for an atom at rest. The red shift of the $m_F=+1$ component is principally due to the second-order Doppler effect. On the contrary, because of the motional Stark effect, the $m_F=-1$ component is blue shifted and broadened and its intensity is reduced.}
\label{fig:10}       
\end{figure}

For an atom in the initial state $g$ with the velocity $v$ in the magnetic field $B$, the detected fluorescence $F_g(\omega, v, B)$ is:
\begin{equation}
F_g(\omega, v, B)=D\left(\Gamma_e \sigma_{ee}+\sum\limits_{f}\gamma_f \Gamma_f \sigma_{ff}\right)
\label{Fluorescence}
\end{equation}
where $D$ describes the efficiency of the photon detection and $\gamma_f$ is the branching ratio of the fluorescence from the 3P levels towards the 2S levels (about 0.1183). The populations are calculated following equation (\ref{Population7}) and, to take into account the second-order Doppler effect, the laser frequency $\omega$ is replaced by $\omega(1+v^2/2c^2)$ in the expression of $\Delta_e$ and $\Delta_f$. Figure \ref{fig:10} shows the line shape of the two components $m_F=1$ and $m_F=-1$ for a velocity of 3 km/s and a magnetic field of 17.1 mT. This calculation takes into account the Stark mixing of the 3S$_{1/2}$($F=1,m_{F}=\pm1$) with all the 3P$_{1/2}$ and 3P$_{3/2}$ levels. For the $m_F=1$ component, the effect of the Stark mixing is negligible and the line is red shifted by the second-order Doppler effect. On the other hand, the $m_F=-1$ component is blue shifted and the shift due to the Stark mixing is larger than the second-order Doppler effect. The widths of the two components are also different: 1.451 MHz for the $m_F=-1$ component and 1.006 MHz for the $m_F=1$ one (the natural width of the 3S level is 1.005 MHz).

In the precedent calculation, we have considered an atom which is continuously inside the UV beam and equation (\ref{Fluorescence}) gives the probability per unit of time that we detect a 656 nm photon. Now we have to take into account the velocity distribution of the atoms. The atomic velocity distribution of a hydrogen beam produced by a radio frequency discharge dissociator has been studied by Jaduszliwer and Chan \cite{Jet}. In our experimental conditions (dissociator pressure in the range of 0.4~torr), we
can assume that this distribution $f(v, \sigma)$ is close to the Maxwellian form:
\begin{equation}
f(v,\sigma)=v^{3}\exp(-v^{2}/2\sigma ^{2})
\label{vitesse}
\end{equation}
where $\sigma =\sqrt{kT/M}$ ($T$ temperature, $M$ atomic mass). Then the line shape $R(\omega, \sigma, B)$ is:
\begin{equation}
R(\omega, \sigma, B)=a\int _{0}^{\infty }\frac{1}{v}f(v,\sigma)\sum\limits_{g}F_g(\omega, v, B)dv
\label{forme1}
\end{equation}
where $g$ are the levels 3S$_{1/2}$($F=1,m_{F}=\pm1)$ and $a$ a normalization factor which depends of the atomic density in the atomic beam. If $L$ is the length of the part of the atomic beam in front of the detection system, the two-photon excitation probability is proportional to the  transit time $L/v$. Consequently, we have introduced the factor $1/v$ into equation (\ref{forme1}). In our calculation, we have also taken into account a slight dependence of the two-photon excitation probability with the velocity $v$ which is due to the form of the UV gaussian beam and the life time $\tau$ of the excited level: if the detection is made exactly at the waist of the UV beam, the excitation has been made at a distance $v\tau$ from this waist. Then, in equation (\ref{Rabi2ph}), the power density of the light $I$ depends on the velocity $v$. The calculation of this effect is described in detail in the reference \cite{TheseArnoult}.

\begin{figure}
\resizebox{1.0\columnwidth}{!}{%
  \includegraphics{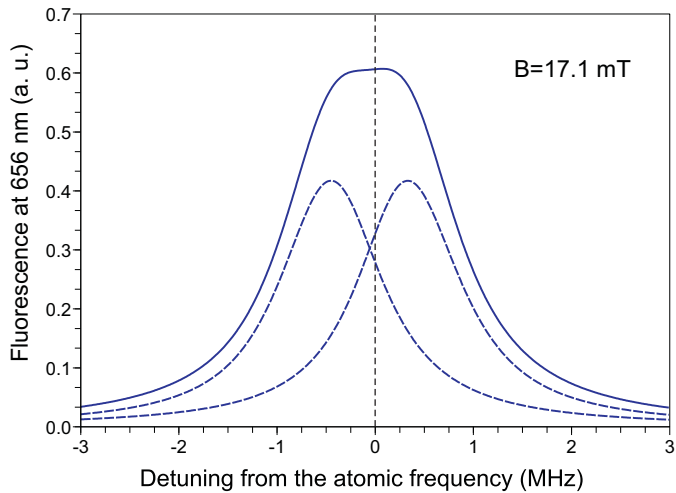}
}
\caption{Theoretical line shape of the 1S$_{1/2}$($F=1,m_F=\pm 1$)-3S$_{1/2}$($F=1,m_F=\pm 1$) transition for a thermal atomic beam ($\sigma = 1.6$ km/s) in a magnetic field of 17.1 mT. The broadening and the shift due to the modulation of the length of the BBO cavity are taken into account with the parameter $\Delta = 57~ {\rm kHz}$. The full line is the sum of the two contributions due to an increase or a decrease in the cavity length (dashed lines).}
\label{fig:11}       
\end{figure}

As explained in section \ref{sec:21}, the length of the BBO cavity is overmodulated: the UV intensity consists in a succession of pulses which corresponds to an increase or a decrease in the cavity length. A splitting and a broadening of the line consequently appear. This effect is described in reference \cite{DeuxBosses}. Following \cite{FORME79}, the line shape of the two-photon transition is given by:
\begin{equation}
\int_{-\infty }^{+\infty }\left| F(\Omega )\right| ^{2}R\left(\frac{\Omega}{2}, \sigma, B\right)d\Omega
\label{convolution}
\end{equation}
where $F(\Omega )$ is the Fourier transform of the square of the electric field at 205~nm.

If $\omega_L/2\pi$ is the laser frequency at 410 nm inside the BBO cavity, the electric field is:
\begin{equation}
E_{410}(t) =\mathop{Re}\left[ \frac{\tau_C E_{0}\exp( i\omega _{L}t)}{1-r_C\exp( i\delta t)}\right]
\label{E410}
\end{equation}
where $E_{0}e^{i\omega _{L}t}$ is the incident field on the BBO cavity, $\tau_C $ the transmission of the input mirror, $r_C$ the reflection coefficient for a round trip in the cavity and $\delta /2\pi $ the frequency shift due to the motion of the mirror ($\delta >0$ if the cavity length decreases). The equation (\ref{E410}) is valid because the lifetime of a photon inside the BBO cavity (about 0.2 $\mu$s) is short in comparison with the duration of the UV pulse (6 $\mu$s). Then the electric field at 205 nm generated by the BBO crystal is proportional to:
\begin{equation}
E_{205}(t)\sim\mathop{Re}\left[ \frac{\tau_C E_{0}\exp( i\omega _{L}t)}{1-r_C\exp( i\delta t)}\right] ^{2}
\label{E205}
\end{equation}
Then the function $F(\Omega )$ is:
\begin{equation}
F(\Omega )=\frac{1}{2\pi }\int_{-\infty }^{+\infty }\frac{\tau_C
^{4}E_{0}^{4}\exp \left[ i\left( \Omega -4\omega _{L}\right) t\right] }{%
4\left[ 1-r_C\exp (-i\delta t)\right] ^{4}}dt{\rm ,}  \label{Fourier}
\end{equation}
where we have only kept the term resonant with the two-photon transition. As the duration of the UV pulses (6 $\mu$s) is short with respect to the period of the modulation (67 $\mu$s), we can linearize the denominator in equation (\ref{Fourier}) and obtain $\left| F(\Omega )\right| ^{2}$. When the length of the cavity decreases ($\delta >0$), we have:
\begin{eqnarray}
\left| F(\Omega )\right| ^{2} &=&0{\rm \qquad if\qquad \ }\Omega <4\omega_{L}{\rm ,}  \nonumber \\
\left| F(\Omega )\right| ^{2} &=&\left( \frac{\tau_C ^{4}E_{0}^{4}}{24r_C^{4}\delta ^{4}}\right) ^{2}
\left( \Omega -4\omega _{L}\right)^{6} \nonumber \\
&&\times \exp \left[ -\frac{\Omega -4\omega _{L}}{\Delta }\right] \qquad {\rm if}\quad \ \Omega >4\omega _{L}
\label{profil}
\end{eqnarray}
where $\Delta =r_C\delta /2(1-r_C)$ characterizes the shift and the broadening of the line. When the cavity length increases, the function $\left| F(\Omega )\right|^{2} $ is symmetric with respect to $\Omega =4\omega _{L}$.
This symmetry supposes that the frequency shifts $\delta_{incr}$ and $\delta_{decr}$ are the same for the forward and backward scanning. In fact, a non-linear refractive index of the BBO crystal can produce a dissymmetry. This effect will be estimated in section \ref{sec:5}.
From equations (\ref{profil}), a straightforward calculation shows that the mean position of each function $\left| F(\Omega )\right| ^{2}$  is shifted by $\pm 7\Delta $ with respect to $4\omega _{L}$ \cite{These Hagel}.

Finally, the line shape is obtained from equations (\ref{convolution}) and (\ref{profil}). Figure \ref{fig:11} shows the profile calculated with a velocity distribution parameter $\sigma=1.6$ km/s, in a magnetic field of 17.1 mT and for a modulation $\Delta = 57~ {\rm kHz}$. This modulation splits the line shape in two contributions corresponding to an increase or a decrease in the cavity length (dashed lines on figure \ref{fig:11}) with a separation of about 800 kHz.

\subsection{Fit of the experimental profiles}
\label{sec:43}

\begin{figure}
\resizebox{1.0\columnwidth}{!}{%
  \includegraphics{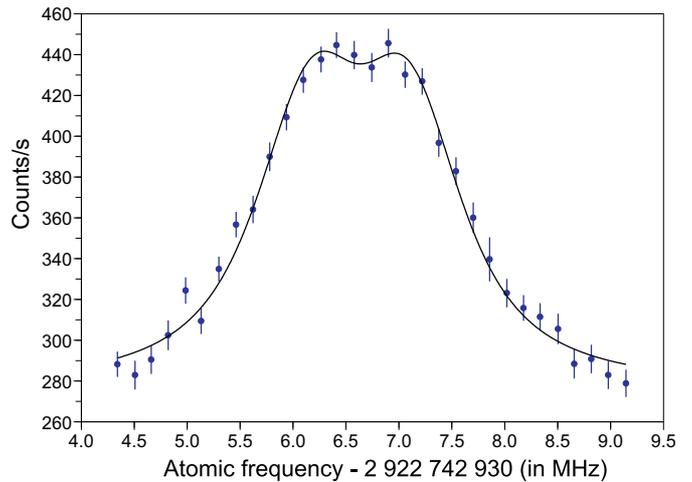}
}
\caption{Fit of the theoretical curve to the experimental data of figure \ref{fig:4}. The velocity distribution width is fixed at $\sigma=1.6$ km/s. The fitted parameters are a modulation $\Delta$ of 69 kHz and a frequency of the 1S-3S transition of $2~922~742~936.745(17)~{\rm MHz}$.
}
\label{fig:12}       
\end{figure}
\begin{figure}
\resizebox{1.0\columnwidth}{!}{%
  \includegraphics{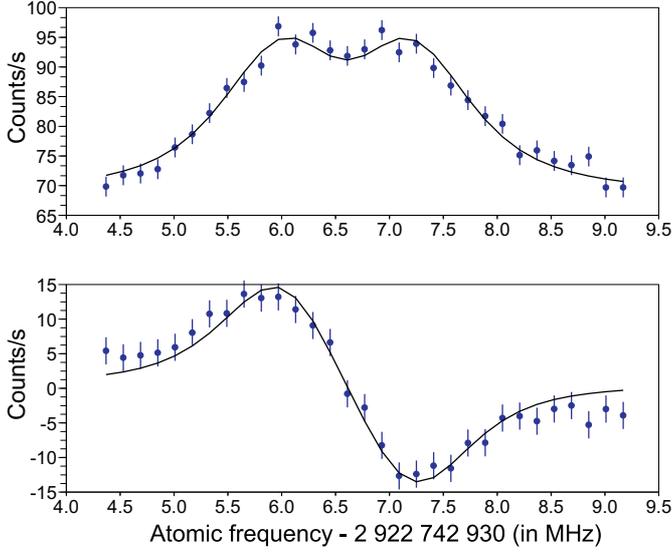}
}
\caption{Fit of the theoretical curves to the experimental data of figure \ref{fig:3}. The upper part of the figure (respectively the lower part) shows the fit of the sum (respectively the difference) of the two curves of figure \ref{fig:3} corresponding to an increase or a decrease in the length of the BBO cavity. The velocity distribution is fixed at $\sigma=1.6$ km/s. The fitted parameters are a modulation $\Delta$ of 92 kHz and a frequency of the 1S-3S transition of $2~922~742~936.722(18)~{\rm MHz}$.
}
\label{fig:13}       
\end{figure}

The parameters of the theoretical curves are the frequency of the 1S-3S transition $\nu _{eg}=(\omega _{e}-\omega _{g})/\hbar$ for an atom at rest without magnetic field, the parameter $\Delta$ which characterizes the splitting due to the modulation of the length of the BBO cavity, and $\sigma$ which describes the velocity distribution. In a simple model the second-order Doppler shift is given by $-3/2(\sigma/c)^2 \nu _{eg}$ \cite{EPJD00}. Consequently the parameters $\nu _{eg}$ and $\sigma$ are strongly correlated and it is not possible to deduce these two parameters from the fit of the line shape. Practically, $\sigma$ is fixed and the fit gives $\Delta$ and $\nu_{eg}$; $\sigma$ will be deduced in the next section from the comparison of the results for different values of the magnetic field.

Figure \ref{fig:12} shows an example of the fit of the theoretical curve to the data obtained with the CCD camera. In addition to $\Delta$ and $\nu_{eg}$, the other parameters of the fit are the amplitude of the signal and an offset corresponding to various detection noises (parasitic light and read-out noise of the CCD camera). The signal-to-noise ratio of this record is about 24 and the atomic frequency is determined with a statistical uncertainty of about 17 kHz.

Through detection with the photomultiplier, it is possible to fit simultaneously the two spectra corresponding to an increase or a decrease in the BBO cavity length. Nevertheless it is preferable to fit the sum and the difference of the two curves, because, in the case of the difference, a large part of the noise due to the UV parasitic light is rejected. An example is shown on figure \ref{fig:13}. The signal-to-noise ratio is about 16 for the fit of the sum of the curves and 13 for the difference. The statistical uncertainty of the atomic frequency $\nu_{eg}$ is 18 kHz.

\section{Results}
\label{sec:5}

\begin{table*}
\caption{Features of the recordings for the different values of the magnetic field.}
\label{tab:1}       
\begin{tabular}{lrrrr}
\hline\noalign{\smallskip}
 Magnetic field & 0.029 mT & 16.01 mT & 17.12 mT & 19.15 mT  \\
\noalign{\smallskip}\hline\noalign{\smallskip}
Number of recordings & 39 & 19 & 66 & 28 \\
Mean frequency $\nu_{eg}-2~922~742~930$ (MHz) &6.724(5)  &6.728(10) &6.721(6) & 6.707(10) \\
$\chi^2/(n-1)$ & 0.69 & 1.03 & 1.21 & 0.56\\
\noalign{\smallskip}\hline
\end{tabular}
\end{table*}

The main results have been obtained by using the CCD camera. The two-photon transition has been observed for four values of the magnetic field, 0.029 mT, 16.01 mT, 17.12 mT and 19.15 mT. The first value corresponds to the residual earth magnetic field. The magnetic field is calibrated with an accuracy of about 0.01 mT by observing the Zeeman effect of the 1S$_{1/2}$-3S$_{1/2}(F=1,m_F=0)$ transition. Table \ref{tab:1} gives, for each value of the magnetic field, the number of recordings $n$, the mean $\nu_{eg}$ of the atomic frequencies obtained by fitting each recording with $\sigma=1.6~{\rm km/s}$ and the values of $\chi^2/(n-1)$. For $B=0.029~{\rm mT}$, the uncertainty of the frequency measurement is 5~kHz, {\it i.e.} a relative uncertainty of $1.7 \times 10^{-12}$. Unfortunately it is not the final accuracy, because the parameter $\sigma$ of the velocity distribution has been fixed to $1.6~{\rm km/s}$.

\begin{figure}
\resizebox{1.0\columnwidth}{!}{%
  \includegraphics{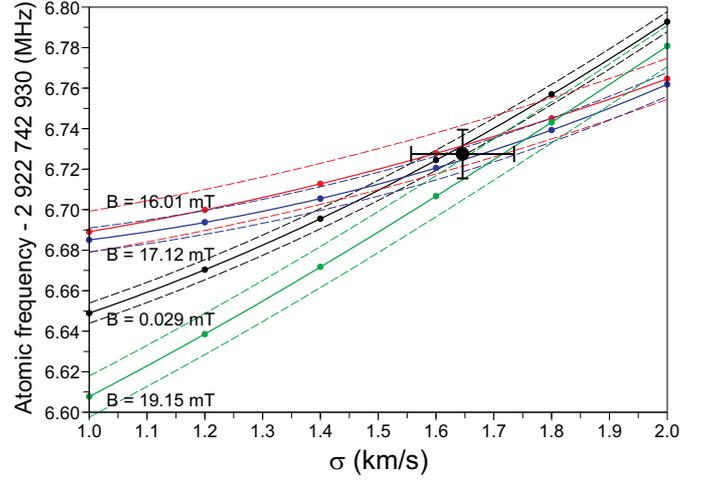}
}
\caption{Frequencies $\nu_{eg}$ as a function of the parameter $\sigma$ for four values of the magnetic field. The calculation has been made for 6 values of $\sigma$ spanning between 1 km/s and 2 km/s (small dots). The full lines are the interpolations between these points and the dashed lines the error bars. The large dot is the best value for $\sigma$ and $\nu_{eg}$ with corresponding uncertainties (color online).
}
\label{fig:14}       
\end{figure}

To determine $\sigma$ the mean frequency $\nu_{eg}$ is calculated for different values of $\sigma$ (in the range 1 to 2~km/s) to obtain four curves $\nu_{eg}\left(\sigma,B\right)$ (see figure \ref{fig:14}). The values of $\sigma$ and $\nu_{eg}$ are obtained from the crossing of these four curves. With a least squares method, we obtain:
\begin{equation}
\sigma=1.646~(89)~{\rm km/s}
\label{sigma}
\end{equation}
\begin{equation}
\nu_{eg}=2~922~742~936.7275~(120)~{\rm MHz}
\label{frequence}
\end{equation}
The corresponding point is indicated in figure \ref{fig:14} (large dot); $\nu_{eg}$ is the frequency of the 1S$_{1/2}$-3S$_{1/2}(F=1)$ two-photon transition for an atom at rest. The statistical uncertainty (12 kHz) corresponds to a relative uncertainty of $4.1 \times 10^{-12}$.

For a thermal atomic beam one has $\sigma =\sqrt{kT/M}$ (see section \ref{sec:42}). The value of $\sigma$ (equation (\ref{sigma})) corresponds to a temperature of 328(35)~K which is compatible with a hydrogen atomic beam produced with a radio frequency discharge. This $\sigma$ value is also in agreement with our previous measurements \cite{Hagel}. In this work, we did not measure the absolute frequency of the 1S-3S transition. Nevertheless, by comparing measurements with and without magnetic field, we had observed the effect of the motional electric field and deduced a value for $\sigma$ of 1.55(11)~km/s. For this analysis, we had approximated the theoretical line shape by a Lorentzian shape. We have remade a complete analysis of these data following the procedure described in section \ref{sec:4}. The result is $\sigma=1.633(180)~{\rm km/s}$. This value is more reliable than the result given in reference  \cite{Hagel} and in good agreement with the new result (equation (\ref{sigma})). It is also possible to compare this result with the measurement of the velocity distribution of the metastable 2S atomic beam used to observe the 2S-$n$S and 2S-$n$D two-photon transition in hydrogen \cite{EPJD00}. In these experiments, the metastable atomic beam was obtained by electronic excitation of a 1S atomic beam which had the same design than the atomic beam used in the present experiment. Because of the electron impact, there was an angle $\theta=20{^{\circ }}$ between the 1S and the 2S atomic beams and the metastable atoms were slowed with respect to the 1S atomic beam by a factor of about $cos(\theta)$. Using Doppler spectroscopy of the Balmer-$\alpha$ line, the $\sigma$ value of the metastable atomic beam was 1.525(10)~km/s \cite{JPhys2}. Taking into account the factor $cos(\theta)$, this corresponds to a value of 1.623(11)~km/s. In spite of the roughness of this model, this value is also in perfect agreement with the present result.

\begin{table}
\caption{Error budget.}
\label{tab:2}       
\begin{tabular}{lr}
\hline\noalign{\smallskip}
Frequency measurements &  $8\times 0.3~{\rm kHz}$ \\
Light shift & $0.3~{\rm kHz}$ \\
Pressure shift   & $1.2~{\rm kHz}$\\
Velocity distribution & $3.0~{\rm kHz}$\\
Scan of the BBO cavity&$2.6~{\rm kHz}$\\
Statistic  & $12.0~{\rm kHz}$\\
\noalign{\smallskip}\hline\noalign{\smallskip}
Quadratic sum  & $13.0~{\rm kHz}$\\
\noalign{\smallskip}\hline
\end{tabular}
\end{table}

The error budget is summarized in Table \ref{tab:2}. The frequency measurements are made with respect to the 100~MHz reference signal from LNE-SYRTE. During the measurements of the 1S-3S transition, this frequency reference was obtained from a hydrogen maser numbered 40805 of LNE-SYRTE. Simultaneously to the 1S-3S measurement this standard was also compared to UTC-OP (Coordinated Universal Time-Observatoire de Paris) and linked to the SI (Syst\`{e}me International). The result was:
\begin{equation}
\nu_{SYRTE}=\nu_{SI}\left[1+\left(575 \pm 6 \right)\times 10^{-15}\right]
\label{freqsyrte}
\end{equation}
That produces a correction of $+1680(18)~{\rm Hz}$ of the 1S-3S frequency. On the other hand, the Allan variance of the frequency measurement of the titanium sapphire laser (about $5\times 10^{-13}$ in 1000 s \cite{TheseArnoult}) corresponds to a statistical uncertainty of 180~Hz for a frequency measurement during a 20 minute run, conservatively rounded to 300~Hz in table \ref{tab:2}.

Following the notations of the reference \cite{JPhys2}, the light shift $\Delta \nu_{ls}$ is:
\begin{equation}
\Delta \nu_{ls}=\left(\frac{2a_0^2}{mc^2\alpha}\right)\left(\beta_e-\beta_g\right){\rm ,}
\label{lightshift}
\end{equation}
where $\beta_g$ and $\beta_e$ are the matrix elements of the light shift operators in atomic units for the level $g$ and $e$. Their values are: $\beta_{1S}=-6.44539$ and $\beta_{3S}=20.9264$ \cite{Delande}. The UV power inside the atomic beam cavity is estimated at the maximum to 2~mW. This produces a light shift of 280~Hz, rounded to 300~Hz in table \ref{tab:2}.

 The pressure in the vacuum apparatus is about $6\times 10^{-5}$ torr. The collision shift of the 1S-3S transition is not known. Nevertheless it would be similar to the ones of the 2S-3P hydrogen transition because these shifts are dominated by the two upper levels of these transitions of which the wave functions have the same spatial size. The shift of the 2S-3P transition in He buffer gas has been measured by Weber {\it et al.} to be about $-9~{\rm MHz/torr}$ \cite{Weber}. Then, to estimate the uncertainty due to the pressure shift, we suppose an upper limit of $20~{\rm MHz/torr}$, corresponding to the value of $1.2~{\rm kHz}$ in table \ref{tab:2}.

 Several other effects can modify the frequency determination. In the theoretical analysis, we suppose that the two Zeeman sub levels 1S$_{1/2}(F=1,m_F=\pm1)$ are equally populated. Actually, for a magnetic field of 20~mT, the energy splitting between these two levels is about $10^{-4}$ times the thermal energy. That produces a small population difference and a frequency shift of about 4~Hz which is negligible. To test the sensitivity of the result to the form of the velocity distribution, an analysis of the data has been made with a different velocity distribution $f(v,\sigma)=v^{4}\exp(-v^{2}/2\sigma ^{2})$. The result of equation (\ref{frequence}) is then shifted by $-3~{\rm kHz}$. We have adopted this value as the upper limit for the uncertainty due to the velocity distribution. A last possible effect is a dissymmetry in the modulation of the BBO cavity, if the velocity of the mirror is not the same when the cavity length decreases or increases. The two Doppler shifts $\delta_{decr}$ and $\delta_{incr}$ due to the motion of the mirror cavity would not be exactly opposite. The maximum relative difference between these shifts is estimated to $2\%$. The analysis of the data with this dissymmetry shifts the frequency of the 1S-3S transition by $2.6~{\rm kHz}$. This value corresponds to the uncertainty labeled "Scan of the BBO cavity" in Table \ref{tab:2}.

 Taking into account the statistical uncertainty ($12~{\rm kHz}$), the final result is:
\begin{equation}
\nu\left[{\rm 1S-3S}(F=1)\right]=2~922~742~936.729~(13)~{\rm MHz}
\label{frequencefinal}
\end{equation}
The total uncertainty is $13~{\rm kHz}$, {\it i.e.} a relative uncertainty of $4.5\times 10^{-12}$. After the measurement of the 1S-2S transition, this result is the most precise value of an optical frequency in hydrogen. The hyperfine structure of the 1S level is well known \cite{shf1S} and that of the 3S level is evaluated from the Fermi formula with the Breit correction to be $52~609.4~{\rm kHz}$. After correction of the hyperfine splitting, the frequency of the 1S-3S transition is:
\begin{equation}
\nu\left({\rm 1S-3S}\right)=2~922~743~278.678~(13)~{\rm MHz}
\label{frequencecorr}
\end{equation}

Unfortunately, the accuracy of this result is not sufficient to improve the determination of the Rydberg constant. Several methods can be used to extract $R_{\infty}$. The first one is to use the measurement of the charge distribution of the proton from electron scattering to calculate the Lamb shifts of the 1S and 3S levels. One then obtains for the Rydberg constant:
\begin{equation}
R_{\infty}=10~973~731.568~75~(18)~{\rm m}^{-1}
\label{rydberg1}
\end{equation}
For the calculation of the Lamb shift, we have taken into account all the terms given in the CODATA report \cite{codata06}. The fundamental constants used are also the values of CODATA, except for the fine structure constant. For that we use the recent result of Gabrielse \cite{Gabrielse2008}. With a relative uncertainty of $1.7\times 10^{-11}$, this result is in acceptable agreement with the value given by the CODATA adjustment of the fundamental constants: $R_{\infty}=10~973~731.568~527~(73)~{\rm m}^{-1}$.

As explained in section \ref{intro}, another method to obtain the Rydberg constant is to compare the 1S-2S and 1S-3S frequencies by using the scaling law of the Lamb shifts \cite{{Karshenboim},{Pachucki}}. The result is:
\begin{equation}
R_{\infty}=10~973~731.568~88~(68)~{\rm m}^{-1}
\label{rydberg2}
\end{equation}
The relative uncertainty of this result is $6.2 \times 10^{-11}$ and it is about ten times less precise than the CODATA value. The reason is that the relative accuracy of the 2S-3S transition which appears in this calculation is only $2.8 \times 10^{-11}$. With this method, one also obtains a value for the radius of the charge distribution of the proton:
\begin{equation}
r_{\rm p}=0.911~(65)~{\rm fm}
\label{proton}
\end{equation}
This value is in agreement with the result deduced from the electron scattering ($r_{\rm p}=0.895~(18)~{\rm fm}$) and from the adjustment of the data in hydrogen and deuterium ($r_{\rm p}=0.8760~(78)~{\rm fm}$).

\section{Conclusion}
\label{sec:6}

To conclude, the optical frequency of the 1S-3S transition has been measured for the first time by using a femtosecond frequency comb with a relative accuracy of $4.5\times 10^{-12}$. It is the best measurement of an optical frequency in hydrogen after the one of the 1S-2S transition. The second-order Doppler effect has been determined from the observation of the motional Stark effect due to a transverse magnetic field. A careful theoretical analysis has been presented to describe the main features of the line shape. In spite of this high accuracy, this result does not improve significantly the determination of the Rydberg constant. For that, an accuracy of few $10^{-13}$ would be useful.

 Presently the experiment is mainly limited by the low intensity of the UV source at 205~nm. To circumvent this difficulty we plan to modify the UV source by replacing the two frequency doubling stages at 820~nm and 410~nm by a frequency sum of a UV source at 266~nm and a titanium sapphire laser at 896~nm. With this scheme, similar to the one used by Bergquist to obtain 194~nm \cite{Bergquist}, the UV power would be increased by a factor of ten and the signal-to-noise significantly improved. Moreover it would be also possible to observe the 1S-4S transition in hydrogen which lies at 194.5~nm.

 The authors thank G. Hagel for his essential contribution to the first step of this experiment and P. Clad\'{e} for fruitful discussions on the manuscript. They thank also O. Acef for the frequency measurements of DL/Rb standard and they are indebted to the {\it Service du Temps} of LNE-SYRTE laboratory for the time reference. This work was partially supported by the {\it Bureau National de M\'{e}trologie} (now {\it Laboratoire National d'Essais}).



%

\end{document}